# Machine Learning and Big Scientific Data

Tony Hey, Keith Butler, Sam Jackson and Jeyarajan Thiyagalingam

Scientific Computing Department

Rutherford Appleton Laboratory

Science and Technology Facilities Council

Didcot OX11 0QX, UK

## Abstract

This paper reviews some of the challenges posed by the huge growth of experimental data generated by the new generation of large-scale experiments at UK national facilities at the Rutherford Appleton Laboratory site at Harwell near Oxford. Such 'Big Scientific Data' comes from the Diamond Light Source and Electron Microscopy Facilities, the ISIS Neutron and Muon Facility, and the UK's Central Laser Facility. Increasingly, scientists are now needing to use advanced machine learning and other AI technologies both to automate parts of the data pipeline and also to help find new scientific discoveries in the analysis of their data. For commercially important applications, such as object recognition, natural language processing and automatic translation, deep learning has made dramatic breakthroughs. Google's DeepMind has now also used deep learning technology to develop their AlphaFold tool to make predictions for protein folding. Remarkably, they have been able to achieve some spectacular results for this specific scientific problem. Can deep learning be similarly transformative for other scientific problems? After a brief review of some initial applications of machine learning at the Rutherford Appleton Laboratory, we focus on challenges and opportunities for AI in advancing materials science. Finally, we discuss the importance of developing some realistic machine learning benchmarks using Big Scientific Data coming from a number of different scientific domains. We conclude with some initial examples of our 'SciML' benchmark suite and of the research challenges these benchmarks will enable.

1. **The Deep Learning Revolution and 'AI for Science'**

It is arguable that the Deep Learning revolution we are now witnessing dates back to the ImageNet database and the AlexNet Deep Learning network [1]. ImageNet was a project that was led by Professor Fei-Fei Li from Stanford University and the database contained over 14 million high-resolution images collected from the Web. The images were labeled by crowd-sourcing human labelers recruited using Amazon's Mechanical Turk. Starting in 2010, an annual competition called the ImageNet Large-Scale Visual Recognition Challenge [2] was held using the database. The competition used a subset of the ImageNet collection with roughly 1000 images in each of 1000 categories. In all, there were roughly 1.2 million training images, 50,000 validation images, and 150,000 testing images. The intent was to provide the computer science community with a focus for evaluating the effectiveness and progress of computer vision systems. A landmark breakthrough in image classification was made in 2012 by Geoffrey Hinton and two of his PhD students, Alex Krizhevsky and Ilya Sutskever. AlexNet, as their neural network implementation became to be called, used a so-called 'Deep Neural Network' consisting of five convolutional layers and three fully connected layers and was implemented using two GPUs. Their paper won the 2012 ImageNet competition and reduced the error rate by an astonishing 10.8% compared to the previous winner [3]. The 2015 competition was won by a team from Microsoft Research using a very deep neural network of over 100 layers and achieved an error rate for object recognition comparable to human error rates [4]. In the words of Geoffrey Hinton, 'Deep Learning is an algorithm which has no theoretical limitations on what it can learn; the more data you give and the more computational time you provide the better it is' [5].

Can such AI algorithms benefit scientific research? Google's DeepMind subsidiary in the UK has brought together physicists, machine learning experts and structural biologists to create a system called 'AlphaFold' [6]. The DeepMind team entered the biennial competition organized by CASP (Critical Assessment of protein Structure Prediction) that assesses the state of the art in three-dimensional protein structure modeling [7]. David Baker, a CASP organizer and developer of the Rosetta protein folding program at the University of Washington in Seattle [8], commented that 'DeepMind's scientists built on two algorithm strategies pioneered by others. First, by comparing vast troves of genomic data on other proteins, AlphaFold was able to better decipher which pairs of amino acids were most likely to wind up close to one another in folded proteins. Second, related comparisons also helped them gauge the most probable distance between neighboring pairs of amino acids and the angles at which they bound to their neighbors. Both approaches do better with the more data they evaluate, which makes them more apt to benefit from machine learning computer algorithms, such as AlphaFold, that solve problems by crunching large data sets' [9]. The predictions of the AlphaFold system were remarkably good and better on average than the other 97 competitors. However, there is still hope for scientists. After the competition David Baker remarked that 'Deep Mind made much better fold level predictions than everybody, including us, using DL on co-evolution data. For problems where there are not many homologous sequences, and for protein structure refinement, I would expect their approach to work less well, as it doesn't have any physical chemistry (they used Rosetta to build their final models from predicted distances)' [10].

In this paper, we make some initial explorations into the application of such Deep Learning approaches applied to scientific data. The Rutherford Appleton Laboratory (RAL), at Harwell near Oxford, hosts several large-scale experimental facilities that now generate large volumes of increasingly complex scientific data. These are the Diamond Synchrotron Light Source and Electron Beam Facility, the ISIS Neutron and Muon Facility and the UK's Central Laser Facility. In addition, the Scientific Computing Department at the Laboratory hosts the UK's Tier 1 Centre for particle physics data from the Large Hadron Collider at CERN and the Natural Environmental Research Council's JASMIN 'Super Data Cluster' that supports their Centre for Environmental Data Analysis. The Scientific Machine Learning (SciML) group at the Lab is now partnering with the Alan Turing Institute, the UK's national institute for data science and artificial intelligence, in their new 'AI for Science' research theme. The SciML group will also be providing the 'PEARL' GPU computing service to Turing researchers on two NVIDIA DGX-2 GPU systems.

After outlining three example applications of machine learning applied to data generated by the RAL Facilities, we discuss the challenges in combining experimental and computational simulation data for progress in materials science. The article concludes with a discussion of progress towards the creation of a scientific machine learning benchmark suite.

## 2. Scientific Machine Learning at the Rutherford Appleton Laboratory: Three Examples

### 2.1 Introduction

Machine learning has the potential to be applied for the enhanced operation and functioning of large-scale big science projects. Our work in this area builds on notable successes from the application of machine learning to analyse and interpret data at national facilities, particularly in the USA. At Argonne National Laboratory machine learning is being used to complement reverse Monte Carlo structure determination from scattering experiments, by applying reinforcement learning [11]. Researchers are X-ray tomography are also applying machine learning to assist with experiment orientation and facilitating better signal to noise ratios in low-dose experiments [12,13] and also using machine learning approaches to correlate diffraction and microscopy techniques to allow for advanced characterisation of phenomena such as lattice vibrations [14]. The Advanced Light Source at Berkeley is using machine learning to automate the collection and analysis of data from micro tomography experiments [15] and is also working with Argonne and the Materials Virtual Laboratory to automate the collection and curation of X-ray spectroscopic data [16]. At Oak Ridge National Laboratory machine learning is being applied to the analysis of electron microscopy data for following materials dynamics, such as perovskite octahedral tilting [17] and silicon migration in graphene [18], while researchers at the lab's spallation neutron source have used autoencoders to build physical models from inelastic neutron scattering experiments on a spin-ice material [19]. Example also exist demonstrating how machine learning can be used for the enhanced operation of large facilities, recently a notable effort showed deep learning with multi-modal data could be used to predict plasma instabilities in large-scale fusion reactors [20].

### 2.2 Diamond Light Source and Cryo-Soft X-ray Tomography (CryoSXT)

The first example is from the Diamond Light Source, an experiment on tomographic biological imaging. Cryo-SXT is a 3D imaging method for the visualisation of cellular ultrastructure and specifically addresses the need for detailed, 3D information on cellular features in thick specimens, such as whole cells, with little or no chemical or mechanical modification [21]. A major bottleneck in the analysis of the 3-D images created by such tomograms is in the segmentation of the images to distinguish between the cell nucleus, cytoplasm and the individual organelles. Because there are few pre-labelled image sets, and the diversity of different cells is very large, it is not possible to straightforwardly use Deep Learning techniques. Instead, Michele Darrow and Mark Basham and their team have used 'shallow' machine learning techniques with some user annotation of a subset of images to speed up the segmentation process. These techniques have been incorporated into their SuRVoS segmentation workbench [22]. Figure 1 gives an indication of this process. The team have also enlisted the help of citizen scientists using the Zooinverse platform for their 'Science Scribbler' project [23].

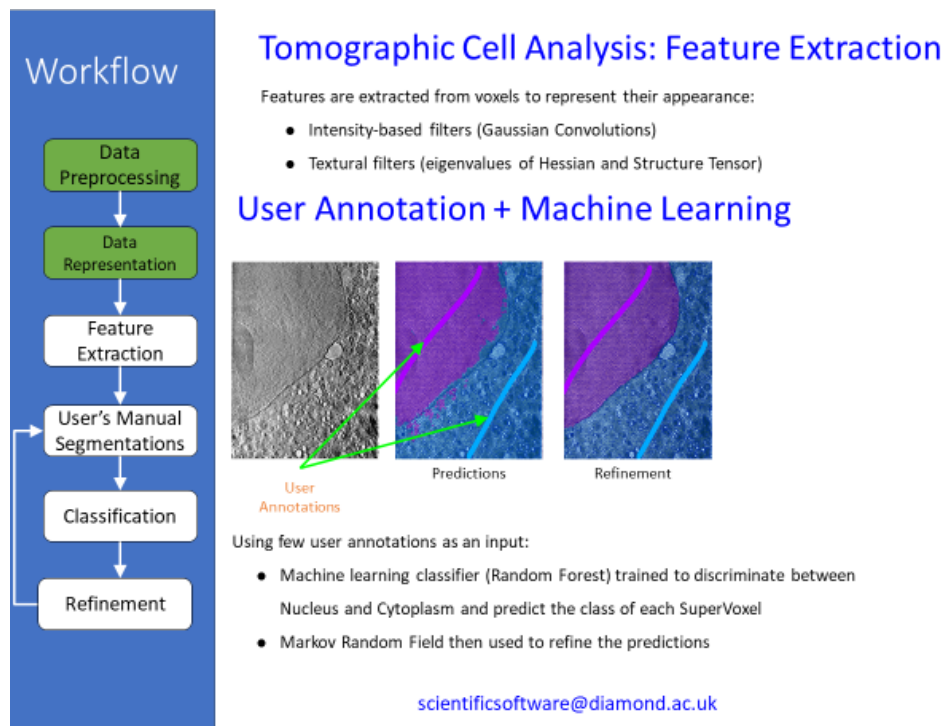

Fig. 1: Schematic representation of the workflow for a Cryo-Soft X-ray tomography experiment showing how user annotation of a few images can be used to train a machine learning classifier to distinguish between the cell nucleus and cytoplasm. (Thanks to Mark Basham, DLS.)

### 2.3 Electron Cryo-Microscopy

Thanks to improvements in instrumentation and software, the use of electron cryo-microscopy (cryoEM) in molecular and cellular biology has increased dramatically in recent years. In the technique of *single particle reconstruction*, micrographs taken from flash-frozen samples of purified macromolecular complexes allow the reconstruction of high-resolution 3D molecular structures from multiple 2D views [24]. Figure 2 shows a schematic of the cryoEM data processing pipeline [25].

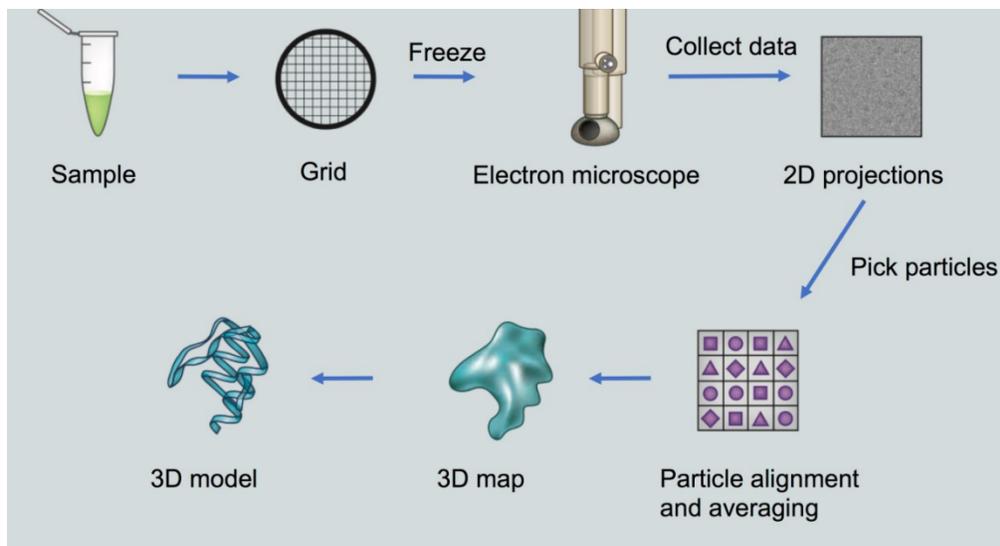

Fig. 2 A schematic of the single particle reconstruction cryoEM pipeline. Image thanks to Creative Biostructure, https://www.creative-biostructure.com

Investment in infrastructure, such as the eBIC facility on the Harwell campus [26] and CCP-EM [27], is supporting a rapid expansion in the use of cryoEM in structural studies. Structure determination involves a sequence of steps, each of which could benefit from algorithmic and computational improvements. Given the large amount of data produced by modern microscopes and detectors, much of which is archived by the facilities themselves or by repositories such as EMPIAR [28], this is a promising area for data-driven approaches.

Machine learning approaches are beginning to be developed for cryoEM, in particular exploiting recent advances in image recognition. The particle picking step of single particle analysis, in which individual molecular complexes are located in micrographs, has attracted the most attention. Due to the intrinsic low contrast of soft matter, and the low dose applied in order to avoid radiation damage, identification of particles is not trivial. Particles represent different views of a molecular complex, depending on their orientation in the sample, and there may be multiple molecular species, as well as contaminants and other artefacts. Furthermore, tens or hundreds of thousands need to be identified from a set of micrographs to yield a high-resolution 3D reconstruction. The promise of machine learning is to reduce the amount of human time required to validate automatically picked particles or to pick missed ones.

As a recent example, Topaz is one of several new programs that use CNNs to learn how to recognise particles in a micrograph [29]. It uses a positive vs unlabelled classification scheme to train a model based on a small set of annotated particles. In the case of a ribosome dataset, Topaz picked 1.72x more particles than the published picks, resulting in the highest resolution structure of this dataset to date.

### 2.4 Fluorescence Localisation Imaging with Photobleaching (FLImP)

The OCTOPUS (Optics Clustered To OutPut Unique Solutions) imaging facility in the Central Laser Facility at RAL combines multidisciplinary expertise, techniques and infrastructure to generate and explore data for understanding biological processes from the scale of cells to single molecules [30]. Automation is increasingly important to help address this challenge, both to increase throughput and exploitation of the instrumentation, and to reduce the expertise needed by users to use the facility and translate its methods outside the facility environment. A key project is a focus on automation of the Fluorescence Localisation Imaging with Photobleaching (FLImP) method. FLImP, developed in OCTOPUS, is a single molecule method which allows molecular structure determination in fixed cells at ~5nm. It has been used to measure structural fingerprints of cancer-causing proteins in cells with unprecedented detail [31].

A recent collaboration led by the OCTOPUS team, and involving partners from medicine, the pharmaceutical industry and a commercial instrumentation company, seeks to utilise deep learning techniques to automate FLImP to deliver a convenient, high-throughput assay for more efficient use of FLImP in the laboratory and to translate it to the clinic as a new method for precise, personalized cancer diagnosis and treatment. As with all super-resolution imaging techniques, FLImP requires images that meet a number of conditions for implementation of the technique, including: the correct density of fluorescently labelled proteins, the ability to differentiate cells from non-specific background labelling, relative background homogeneity and obtaining sufficient frames to observe the required number of photo-bleaching events and photons. At present, successful FLImP imaging is a user-intensive process, requiring operators to manually select regions of interest for image acquisition. The successful translation of FLImP technology from bench to bedside therefore requires the automation of this image segmentation task to enable the scale-up of this technology. This is a challenging problem, particularly as single fluorescently labelled proteins are diffraction limited in size and therefore difficult to individually segment from images using conventional convolutional neural networks [32]. To this end, the OCTOPUS team have utilised a UNET based model capable of automatically and rapidly segmenting regions of appropriate FLImP object density from micrographs derived from mono-cell cultures (Figure 3 (C)). The team is currently working with clinical collaborators to extend this technique to permit multi-label classification to identify cells of interest from more diverse clinically derived samples that may include cells from a number of populations, only some of which are suitable for FLImP, and integrate these models into instrumentation suitable for clinical translation.

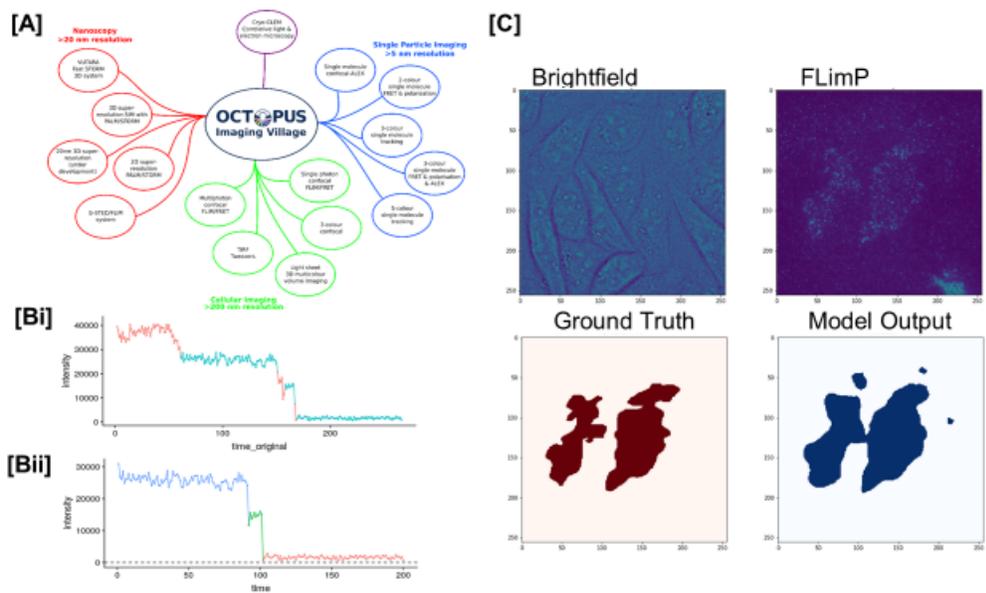

Fig. 3: [A] Overview of the techniques employed at OCTOPUS. [B] An illustration of the automating FLImP integrated intensity track selection process. [Bi] Raw FLImP track showing regions deemed suitable (blue) and unsuitable (red) for FLImP analysis. [Bii] Processed FLImP track from [Bi] with distinct levels detected. [C] Automatic detection of regions suitable for FLImP using a deep learning approach.

## 3. Machine Learning and Materials Science

### 3.1 Overview of Materials Science and Machine Learning

Machine learning has started to change the way that we do materials science; contributing to accelerated characterization, synthesis and modelling. These advances are driven by the availability of easy to use packages for building machine learning models, e.g. Scikit-Learn [33] and Keras [34], as well as a recent proliferation of publicly available datasets, resulting in a materials science "ImageNet moment" where the availability of data fuels a step-change in data-driven approaches. We will briefly survey some of the cutting-edge machine learning work in the areas of materials discovery and characterization as well as outlining some of the work of the SciML team that is using machine learning to analyse data produced at the UK's large national scientific facilities.

Computational materials science dates back to mid-twentieth century, an early example being the quantum chemistry exchange programme, which allowed experimental chemists to perform quantum chemical calculations with relative ease [35]. At this early stage, the paradigm of computational materials science was to use computational methods to help interpret experimental results by doing a small number of expensive calculations on materials whose structure was already well known. Density functional theory (DFT) was popularized by Walter Kohn and co-workers in the 1960s; with the advent of powerful super-computers in the late twentieth century performing large numbers of DFT calculations suddenly became feasible [36, 37]. Structure prediction methods based on global optimization algorithms such as particle swarm optimization and genetic algorithms mean that it is now possible to predict structure and properties for new materials starting from the composition alone [38]. The availability of rapid and accurate DFT calculations has also facilitated the development of large, high-quality databases of calculated materials properties, for example the Materials Project, Aflow, Open Quantum Materials Database and Nomad [39].

The sudden availability of these datasets is revolutionizing the way that data-driven approaches are used in materials science. Figure 4 plots the number of publications containing "machine learning materials" in them from the Web of Science. We also indicate on the figure dates that some notable databases became available, suggestive of the important role of these datasets are playing in driving the development of a new paradigm of computational materials science [40].

New machine learning approaches trained on computational databases are capable of making rapid and accurate predictions of materials properties by considering composition alone. The electronic band gap is a good example of a material property that is important in a range of applications from microelectronics to photovoltaics. A number of studies have reported machine learning algorithms that are capable of predicting the band gap of a material from its composition [41 - 43]. These kinds of algorithms can be incorporated into materials discovery workflows and have recently been applied to the prediction of new photoactive earth-abundant materials for photocatalysis [44]. Generative models, using neural networks, are also now being used to postulate new molecular materials [45]. For example, the long short-term memory (LSTM) neural network architecture has recently been shown to be able to predict new drug molecules using greatly reduced training data compared to other approaches [46]. In the ORGAN project, a combination of generative adversarial network and reinforcement learning combine to bias molecular generation algorithms towards desired final metrics, potentially allowing the automated design of a molecule to meet a specific property [47].

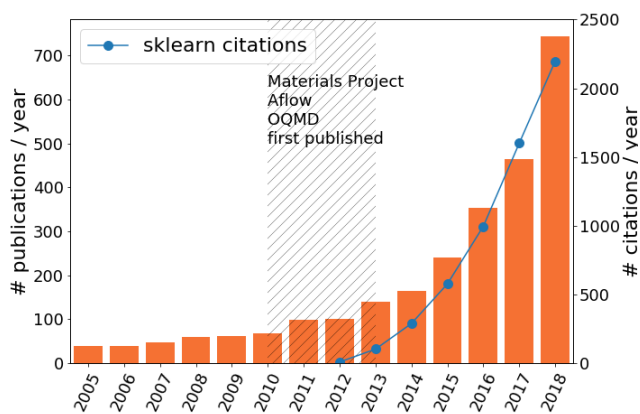

Figure 4: The ML explosion in materials science. The number of papers containing the terms machine learning and materials are plotted on a bar chart. We also indicate the dates of materials data repositories becoming available and plot the number of citations for popular machine learning toolkit, Scikit-Learn over the same period.

Interpretation of complicated experimental spectra has regularly relied on calculations for clarification, but now with databases of calculated properties available it is possible to develop machine learning algorithms to interpret spectra in an automated way, free from human bias and capable of identifying signals which are missed during manual inspection. A powerful recent example is in the field of X-ray absorption spectroscopy (XAS) where a dataset of calculated spectra was recently made available [48]. Calculated spectra have been used to train neural networks, which are facilitating unprecedented analysis of materials datasets, for example, in characterising structural transformations in materials, in making on-the-fly predictions about the presence of chemical environments in a sample and in identifying sub-nanometer atomic assemblies [49 – 51]. Recently an ensemble learning algorithm trained on this dataset that is capable of identifying the oxidation state and coordination environment in a diverse range of chemistries has been made publicly available [52].

### 3.2 Machine Learning and Experimental Materials Data

The rapidly expanding capability of large-scale facilities to analyse material samples means that the demand for robust, automated, on-the-fly analysis is becoming ever more pressing. Examples such as the XAS studies described above show how a fusion of experiment, simulated data and machine learning algorithms can facilitate rapid interpretation of these rich new data sources. In the SciML team we are developing a range of machine learning algorithms for materials data analysis.

Inspired and challenged by the progress in machine learning at other large scale facilities outlined in the start of section 2 we have started to build a machine learning capability at the Rutherford Appleton Laboratory for analysis of materials science data collected on site. Here we present our work on diffuse multiple scattering experiments at the Diamond Light Source and on inelastic neutron scattering experiments at ISIS neutron and muon source.

**Diffuse multiple scattering**

Diffuse multiple scattering (DMS) is a relatively new crystallographic technique and has been made possible by the immense increase in flux of modern synchrotron sources and modern detector systems [53]. DMS can be a powerful technique for allowing measurement of fine details such as lattice strain and for following structural phase transitions in materials. However, the detailed experimental setup requires expert knowledge and several time-consuming steps which limit the routine application of the technique.

One of the parameters that must be known for experimental analysis of DMS data is the azimuthal angle of the sample, which is not known *a priori* and determines the values at which reciprocal crystal lattice vectors cross the Ewald sphere, as defined in [53]. We have trained a neural network consisting of convolutional and densely connected layers to predict the azimuthal angle of sample based on the observed scattering pattern. Typically, determination of the azimuthal angle is time-consuming task, requiring expert knowledge and representing a serious bottleneck for application of DMS.

We have built a database of 250,000+ simulated patterns, **Ψ(R)$_T$**, using the DMS Python code, which are used to train the neural network [53]. The simulated patterns provide a labelled ground truth of azimuthal angles, as a function of the patterns **Ψ(R)$_T$**. We then train our NN to predict $\Psi$ based on the input image **R**, updating the filters, weights and biases of the NN to minimize the difference between predicted **Ψ(R)$_{NN}$** and **Ψ(R)$_T$.** The NN that we train is then capable of predicting the azimuthal angle to within 6.5°, see Figure 5. The NN, once trained, can provide an answer in a fraction of the time required for exhaustive comparison of images.

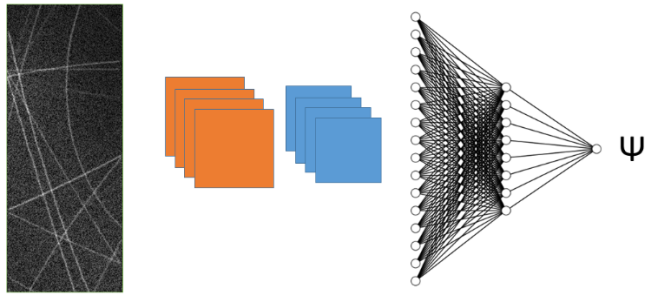

Figure 5  Schematic representation of the CNN used to predict coupling azimuthal angle from diffuse multiple scattering images. A 2D map of multiple scattering lines is passed through 2 convolutional layers, flattened and passed through two densely connected layers and finally passed to a single output node for Ψ. Note that the numbers of filters and nodes are just for illustration, see methods section on DMS network for details.

**DMS network methods:** The NN used for predicting the azimuthal angle of a DMS sample consists of convolutional and densely connected layers. The first convolutional layer contains 32 3x3 kernel filters, followed by a maxpooling of 2x2; the second convolutional layer contains 64 3x3 filters, followed by maxpooling of 2x2. We include a dropout rate of 0.2 between the convolutional layers to guard against over-fitting. The 2D data is then flattened and fed into a densely connected layer of 32 nodes, connected to a densely connected layer of 16 nodes. The final hidden layer is connected to a single output node with a linear activation function to allow the network to perform regression. All hidden layers are connected with rectified linear unit (ReLU) activation functions. The network is trained on 75% of the dataset and then validated on the remaining 25%. The training and validation curves are shown in Figure 6.

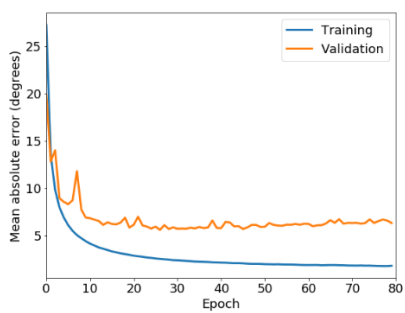

Figure 6 Training and validation scores for the mean absolute error for prediction of the azimuthal angle of a DMS pattern.

**Magnon neutron scattering**

Inelastic neutron scattering can provide detailed information about microscopic materials structure. In particular, the magnetic moment of neutrons allows one to probe the magnetic structure and ordering in a material. In this example we have investigated the use of NNs for predicting the magnetic coupling constants (**J**) in $Rb_2MnF_4$. $Rb_2MnF_4$ is a near-ideal 2D, spin 5/2 Heisenberg antiferromagnet and has been used extensively to test predictions for the 2D Heisenberg quantum Hamiltonian [54, 55]. As such, this system provides an ideal test case for exploring the ability of a NN for this task.

$Rb_2MnF_4$ consists of planes of $MnF_2$ layers, with magnetic Mn arranged in a square lattice. Experimentally it has been established that $Rb_2MnF_4$ has magnetic coupling between nearest neighbor Mn sites with a coupling constant variously measured as J=0.648±0.003, 0.6544±0.014 and 0.673±0.028 meV depending on the experiment and fitting model [54, 55]. Careful examination of the spin wave energies along the antiferromagnetic zone boundary reveal that in addition to the nearest neighbor coupling, there is a next-nearest neighbor term in the Hamiltonian J', which has been measured to be 0.006±0.003 and 0.012±0.002 meV in different experiments [55, 56].

In our study we built a training set of 29957 simulated spin wave spectra in the 2D (*h, k, 0*) plane from 0 ≤ h, k < 1 of the of $Rb_2MnF_4$ using the SpinW code [57]. This serves as our labelled training set **R**. We then train our NN to learn the relation between **R** and **(**J, J'**)**; (J, J') = *f*(**R**), where the function *f* is the NN. After training (details below), we obtain a NN that has a mean average error of ±0.0055 meV on J and ±0.0036 meV on J', using data that was not included in the training set. As a true test of the NN we provided experimental data collected on the MARI instrument at the ISIS neutron and muon source. The data was collected on a sample of $Rb_2MnF_4$ and the image of the integrated energy over the over the plane is shown in Figure 7.

The NN trained on simulated data predicts a value of J=0.6763 meV and J'=0.0104 meV for the experimental spectrum, in excellent agreement with previous experimental results. This demonstrates the ability of a convolutional NN to learn to predict magnetic coupling constants from simulated data, even picking up subtle, difficult to spot features, such as the value of the next-nearest-neighbour coupling constant J'. We stress here that prior knowledge was used to select a training set representative of a reasonable range of final values together with our intuition about the number of coupling constants present. This fusion of prior knowledge and NN architectures helps to greatly improve the efficiency of training and allows development of high-quality models with significantly less data than would otherwise be required. We consider this an example of how NN can be used to augment existing expertise and assist in difficult analysis where some prior knowledge already exists.

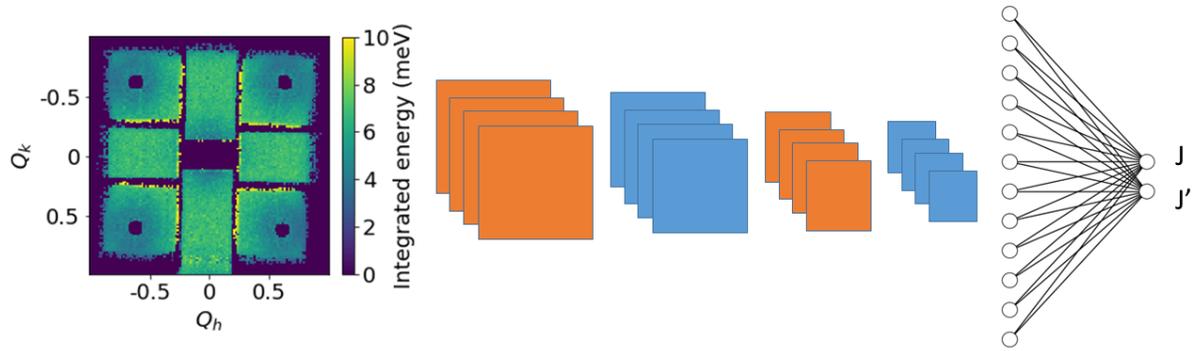

Fig. 7 Schematic representation of the CNN used to predict coupling constants from inelastic neutron scattering images. A 2D map of integrated energy is passed through 4 convolutional layers, flattened and densely connected to two output nodes for J and J'. Note numbers of filters and nodes are just for illustration, see methods section on Magnon network for details.

**Magnon network methods:** The NN used for predicting the magnetic coupling constants consist of four convolutional layers terminated by a densely connected layer. The first convolutional layer contains 32 3x3 kernel filters; the second convolutional layer contains 64 3x3 filters; the third convolutional layer contains 32 3x3 kernel filters; and the final convolutional layer contains 16 3x3 kernel filters, all convolutional layers are followed by maxpooling of 2x2. The 2D data is then flattened and fed into a densely connected layer of 2 nodes with a linear activation function to allow the network to perform regression. All hidden layers are connected with ReLU activation functions.

The network is trained on 27000 images of the dataset and then validated on the remaining 2957 images. The training and validation curves are shown in Figure 8. Before feeding the simulated images into the network they are converted to a 2D histogram of 128 x 128, we also apply a mask to the simulated data to cover the regions of the pattern that are not recorded due to the detector geometry – these appear as areas of purple in the image in Figure 7.

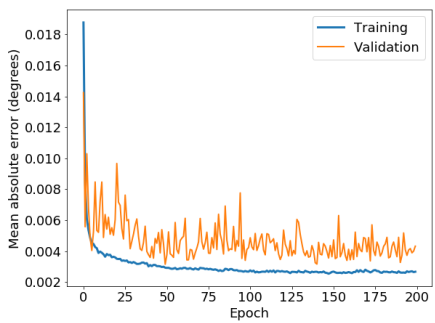

Figure 8 Training and validation scores for the mean absolute error for prediction of the coupling constants from an inelastic neutron scattering pattern.

**3.3 Further work**

In the examples given here, we have used convolutional neural nets (CNNs) to analyse spectra and patterns collected at synchrotron facilities and represented as images. Deep CNNs have revolutionised the field of image processing and recognition in many fields of business and research. As alluded to earlier, the explosion in popularity of NNs, and in particular of deep CNNs for image applications, has been driven largely by the availability of large labelled datasets for training. Deep CNNs typically rely on combinations of many types of operation and connection to achieve their most powerful results on the most difficult problems. This results in networks that not only require vast amounts of labelled data, but also have many tunable hyper-parameters. This can hamper application in many materials science problems, where labelled datasets are limited.

Recently a new type of image recognition architecture, the mixed-scale dense MSD-NN neural network was introduced by researchers at Berkeley Laboratory [58]. This architecture has several differences from traditional CNNs. The MSD-NN uses dilation filters rather than traditional convolutional kernels, this means that longer range correlations in images can be captured, depending on dilation settings (see Figure 9). In the MSD-NN all of the convolved layers are also fully connected, unlike a CNN where layers connect sequentially. This full connectivity means that the network does not have to remember information from layer to layer for the final outcome.

In initial work it has been shown that the MSD-NN can learn on significantly smaller datasets and with less hyper-parameter tuning than CNNs. In SciML, we are currently exploring the application of MSD-NNs for soft X-ray image segmentation and for a range of materials' science classification problems.

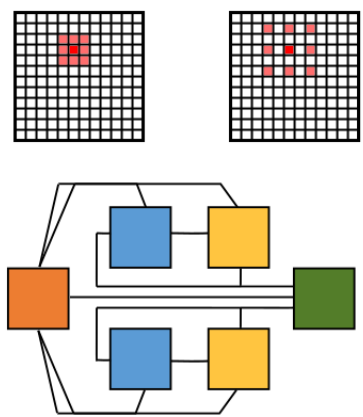

Figure 9 Top an illustration of a typical convolution filter (left) which convolves information from neighbour pixels and a dilation filter (right) which can convolve with pixels further removed. Lower a schematic representation of the fully connected mixed-dense neural network architecture.

## 4. Big Scientific Data and Machine Learning Benchmarks

### 4.1 Introduction

Benchmarking, as a means for measuring and assessing the performance of a given computer system, or as a software framework, has been a cornerstone of computing for years. Historically, these efforts began using small kernels or excerpts of loops, such as the Lawrence Livermore Loops, the Dhrystone and Whetstone benchmarks, and the LINPACK linear algebra benchmark. The key driver for these efforts was to compare runtime performance or the number of floating-point operations per second (Mflop/s) for different computer architectures. Over the years, however, the art of performance evaluation has changed to include a suite of benchmarks, such as the SPEC, ASCI, SPLASH, and NPB benchmark suites, and the measured parameters included multiple metrics, such as runtime performance and energy consumption. Although LINPACK is still a major baseline benchmark for estimating peak performance of the TOP500 supercomputing systems in the world, detailed system evaluation relies on multiple benchmark measurements covering multiple hardware platforms, often incorporating both CPUs and GPUs. More recently, new benchmark suites have been developed to cover other important aspects of computing systems such as storage and networking, with appropriate metrics.

Although architectures and compilers still play a critical role in the development of computing systems, the performance of machine learning systems is now becoming equally important with the rise of commercial applications of machine learning. As can be seen from our discussion above, machine learning is now also beginning to play a major role in scientific applications. Suitable scientific machine learning benchmarks are therefore now required not only to assess systems for these applications but also to assess the overall machine learning ecosystem in a scientific context. The complexity and diversity of machine learning frameworks, available hardware systems, evaluation techniques, suitable metrics for quantification, and the limited availability of appropriate scientific datasets make this a challenging endeavour. Early initiatives on this front include MLPerf [59], AI Benchmark Suites from BenchCouncil [60], CORAL-2 [61], and Deep500 [62, 63].

- The MLPerf benchmark suite currently relies on several common commercially important machine learning-oriented tasks, such as image and object recognition, speech-to-text, sentiment analysis, translation and recommendation applications along with a set of baseline models [59]. However, it is very likely that the suite will incorporate scientific applications. The key metric of the MLPerf suite is speedup relative to a reference implementation. The MLPerf suite also relies on a number of large-scale datasets, covering different application cases within MLPerf. This collection of datasets is also likely to be extended to include scientific cases.

- The international BenchCouncil [http://www.benchcouncil.org] is organizing an AI System and Algorithm competition in 2019 [60]. A number of their benchmark suites, namely, AIBench, HPC AI500, AIoT Bench, Edge AIBench, and BigDataBench, although not primarily focused on scientific applications, could be a useful complement to the SciML benchmarks proposed here. Each of these benchmark suites targets different domain of problem, such as IoTs or Edge computing devices, and includes a number of different types of benchmarks covering micro kernels, components and applications [64-68].

- The CORAL-2 suite includes a ML/DL micro-benchmark suite that captures operations that are fundamental to deep learning and machine learning [61]. These include dense/sparse matrix multiplications, convolutions, recurrent-layers, and one- and two-dimensional Fast and Discrete Fourier Transform kernels (FFTs and DFTs).

- Finally, the Deep500 effort is predominantly focused on techniques for reliably reporting performance of deep learning applications using metrics such as scalability, throughput, communication volume and time-to-solution [62, 63]. This is more focused on methodology (and a corresponding framework) for quantifying and reporting deep learning performance than on any specific application.

One of the key motivations for this work reported in this paper is the lack of a comprehensive machine learning benchmarking initiative for scientific applications, such as particle physics, earth and environmental science, materials, space and life sciences. Such a scientific benchmark suite would facilitate better understanding of machine learning models and their suitability for different operations in a scientific context, rather than being solely oriented on performance.

### 4.2 The SciML Suite – An Overview

Our scientific machine learning benchmark suite, SciML, is intended to span multiple scientific domains, and cover several of the different types of problems arising in each domain. We will therefore provide a number of reasonably large and complex datasets specific to each domain together with one or more baseline models addressing particular domain-specific problems. In addition, the evaluation metrics for the SciML suite go beyond just the simple runtime performance (or speedup). Our goal is to capture the overall performance of a given scientific application by assessing both the training and inference times per sample, as well as the classification accuracy using one or more appropriate metrics. Here, we use classification accuracy, classification loss and F1 score as the relevant metrics. Classification accuracy is the ratio of correctly predicted outcomes to total number of predictions, and thus higher the accuracy, the better the model. The second metric, classification loss, measures the performance of the model by measuring how the predicted outcomes diverge from the actual ones. Finally, the F1 score is the harmonic mean of precision and recall, where the precision is the number of correct positive results divided by the number of all positive results returned by the classifier, and recall is the number of correct positive results divided by the number of all samples that should have been identified as positive. As such, the F1 score is often more useful than the raw classification accuracy when the class distributions are uneven.

The SciML suite will provide the specification of the task plus a reference implementation and can therefore be used to evaluate:

- Different hardware platforms, such as GPUs, TPUs or CPUs
- Different ML-specific frameworks like TensorFlow or PyTorch
- Different implementation of models

The benchmark suite is currently in development and is intended to cover a number of different scientific domains with several problems of varying degrees of difficulty that demand different machine learning techniques. We discuss two of our prototype benchmarks in the subsections that follow.

**Example 1: Small Angle X-Ray Scattering (SAXS)**

**The problem**

Small Angle X-Ray Scattering (SAXS) is one of the benchmarks within the domain of materials science and is particularly relevant to the structure of materials. SAXS helps identify how different materials are structurally organised at the particle level [69, 70]. Here, the term particle means the collective arrangement of several atoms [70]. At this intermediate level of detail, each material can be regarded as being made up of particles of different shapes, such as spheres, rods, ellipsoids and parallelepipeds, and of different sizes, characterised by relevant parameters [71]. When an X-ray beam is sent through a material, particles within the target diffract the incoming X-rays, and the particle sizes, shapes and orientation with respect to the incoming beam determine the resulting diffraction pattern. The distributions of the scattered X-rays are recorded as two-dimensional images. We illustrate an example of different diffraction patterns in Figure 10.

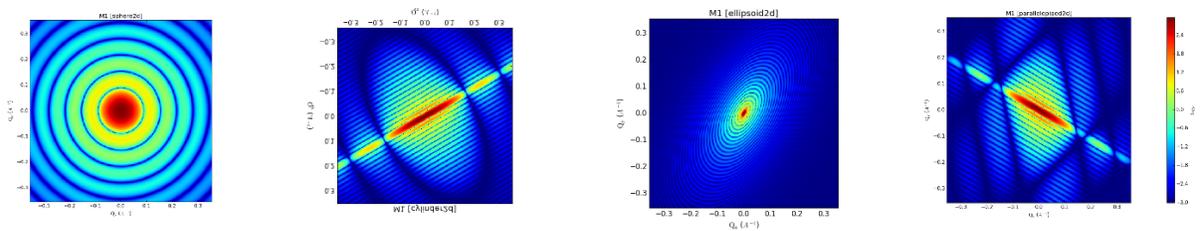

Fig. 10: An example of two-dimensional scattering patterns for sub-shapes of sphere, cylinder, ellipsoid and parallelepiped shapes, from left to right, respectively. The profiles were generated by using the SASView Software [72].

This two-dimensional diffraction pattern, at times, may contain more data than necessary. For instance, in some cases the material can be isotropic with particle arrangements symmetric in every direction, yielding diffraction patterns that are two-dimensionally symmetric. Under these circumstances, a one-dimensional profile can be obtained by integrating the two-dimensional profile over the two-dimensional domain. We show an example for a spherical particle which generates an isotropic two-dimensional profile and the matching one-dimensional profile shown in Figure 11.

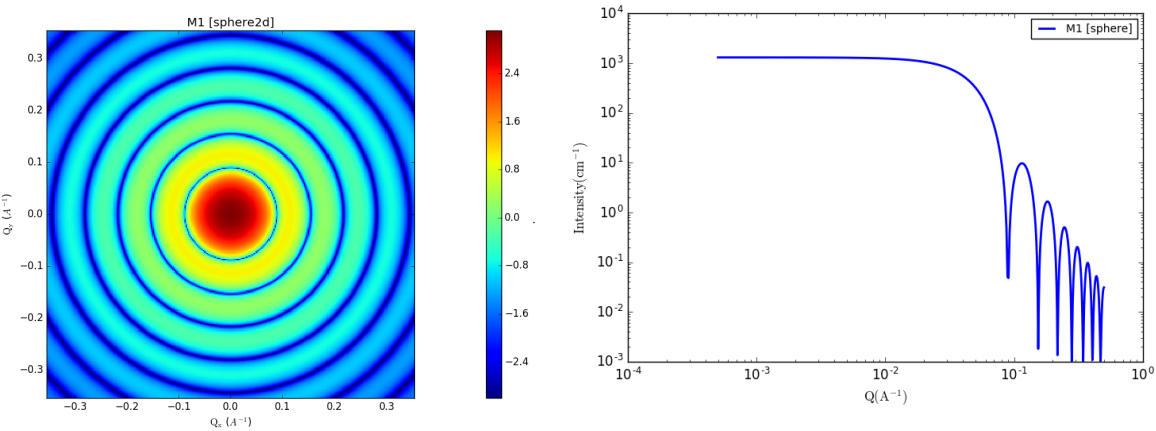

Fig. 11: An example of two-dimensional and one-dimensional scattering profile of a simple spherical particle. The profiles were generated by using the SASView Software [54].

The SAXS benchmark aims to characterise the material given its scattering profile, either one- or two-dimensional. The benchmark uses both simulated and real-world datasets as detailed below. Here, we present a sub-benchmark of SAXS, namely SAXS-1D. This particular sub-benchmark focusses on the binary classification of a set of simulated one-dimensional profiles. The benchmark includes a dataset and a baseline model as discussed below.

**SAXS Dataset**

The SAXS-1D benchmark includes a purely simulated dataset with unit dispersity (particle sizes are not mixed). The actual datasets within the SAXS benchmark are in three categories: ideal simulated datasets; noise-added simulated datasets; and beamline datasets. The first two are generated using the relevant mathematical models [73], while the latter dataset is obtained from one of the beamlines at the Diamond Light Source. The key limitation of the last dataset is that there is no established ground truth, whereas the ground truth information is fully known for the simulated data of the other two cases.

The dataset for this benchmark is focused on identifying two different particle shapes: spheres and parallelepipeds. The sphere is characterised by the radius and is two-dimensionally isotropic. The parallelepiped has three different shape parameters and multiple possible orientations. This sub-benchmark is a simplified case in which the orientation of the parallelepiped and two of the dimensional parameters remain unchanged so that the one-dimensional profile can clearly differentiate the parallelepiped from a sphere.

The simulated dataset contains 10,000 one-dimensional profiles for spheres and 10,000 one-dimensional profiles for parallelepipeds. Out of these, we use 16,000 for training and 4,000 for testing with classes of the particle shapes equally distributed across the datasets. Each of these one-dimensional profiles provides the intensity (I) vs magnitude of the momentum vector (q) and has dimension of [1 x 300].

**Baseline Model**

Although there are several approaches for addressing this challenge, the easiest and perhaps the simplest model is a supervised learning model. Given that the underlying data is obtained through simulation, the ground truth is readily available.

As mentioned in the introductory section, one of the key expectations from the benchmark suite is to obtain a better understanding of different machine learning models and their suitability for different tasks. For this reason, instead of using a more flexible model like a convolutional neural network (CNN) and deep learning, for this sub-benchmark we use a simple, multi-layer neural network for the baseline version. More specifically, this baseline model consists of three densely connected layers, with the first layer capturing the input, which is an array of 300 intensity values, the middle layer with 100 neurons using ReLU as the activation function, and finally the output layer of size one with a sigmoid activation function.

**Example 2: Sentinel Cloud Masking**

This benchmark is intended to capture one of the challenges arising from the earth and atmospheric sciences, namely, the identification of clouds from satellite imagery. This process is often called 'cloud masking'. The masking or quantification of cloud is often an important precursor to using satellite imagery. Clouds are highly dynamic, and this influences their texture, thickness, opaqueness and transparency. The identification process can be very challenging in the presence of snow, sea ice, aerosols and sun glint. We show a cloud masking example in Figure 12.

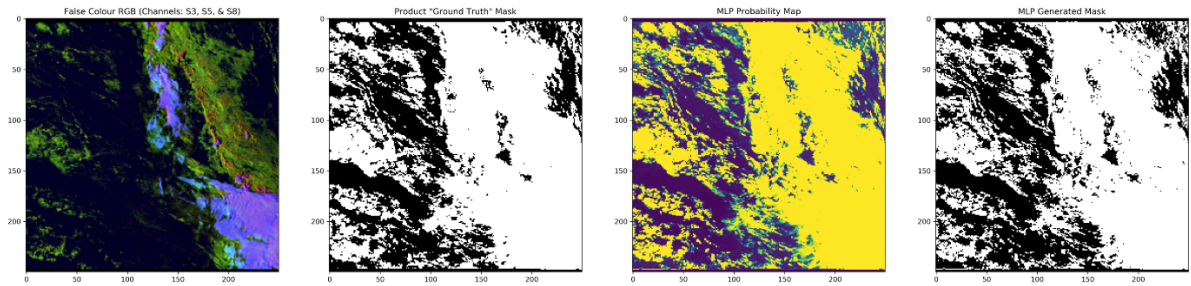

Fig. 12. This shows an example of cloud masking data. Left to right: actual image, ground truth, our generated probability mask, and our generated map. Here, white regions represent the cloud and yellow regions provide the probability map. The colours in the first image are due to the different reflective behavior of different elements in the image, such as sea, ice, land and cloud.

The Sentinel Cloud Masking benchmark will have several sub-benchmarks covering different image modalities, datasets and challenges. Here, we describe the sub-benchmark Sentinel-SLSTR. Sentinel-3 is a constellation of two satellites carrying an array of instruments, including the Sea and Land Surface Radiometer (SLSTR) for measuring sea and land surface temperature, colour and topography to high accuracy. The Sentinel-SLSTR benchmark deals with this specific Sentinel modality. Here we describe the simplest case in which the masking is required above a part of the ocean where there is no sun glint.

The Sentinel-SLSTR benchmark deals with the problem of processing the SLSTR-based data. Given an M x N image, the task is to build a machine learning model for marking each pixel as either cloud or non-cloud, using one of the simplest cases. This benchmark uses the SLSTR images only for the purposes of classification.

**Sentinel SLSTR Dataset**

The overall Sentinel-3 benchmark relies on multiple datasets obtained from different sensors and covers multiple bands in the electromagnetic-spectrum. The Sentinel-SLSTR part of the benchmark uses a collection of 1,000 SLSTR images captured over the South Pacific Ocean region in 2018. The dataset contains significant variation in the number of cloudy pixels with near-ideal illumination of clouds. The data includes 11 channels ranging from very near infra-red, VNIR (0.55 micrometer) to thermal infra-red IR (12.0 micrometer) wavelengths and has two views (nadir and oblique). The spatial resolution is 0.5km in the VNIR and short-wave infra-red (SWIR) channels and 1km in the thermal IR channels. In all experiments the nadir view of channels S1-S9 are used as inputs. To reduce the computational demand, this particular benchmark uses sub-sampled images of 250x250 pixels for each channel. The suite specifies a random selection of 800 images for training with the remaining 200 images for validation.

**Baseline Model**

Unlike our SAXS-1D benchmark that uses simulated data, the key difficulty in building any supervised machine learning model for this Cloud benchmark is the lack of a reliable ground truth. Collective or crowd-sourced hand labelling of these images for ground truth is infeasible for two reasons: the time required for hand-labelling is prohibitive given the volume of images, and secondly, this is a very subjective process even with among experts. For this reason, we use Bayesian inference to generate our surrogate artificial "ground truth" [74-76]. More specifically, for each pixel, we apply the method in reference [74] to mark each pixel as cloud or non-cloud with a corresponding confidence value. This is used as a ground truth in training our networks.

The baseline model we implement for masking cloud on SLSTR data is a plain, multi-layer neural network. Although CNNs or DCNNs have not been used for SLSTR or Sentinel-3 data, many authors have attempted to apply deep learning [77-84] and other complex NN models, such as LSTMs [85] and GANs [86] to cloud screening using other remote sensing instruments. In our case, the neural network-based baseline model consists of three densely connected layers with the first layer capturing the nine-channel images as vectorised inputs, the middle layer with 50 neurons using ReLU as the activation function, and finally the output layer using the sigmoid activation function with one neuron.

**Results**

The SAXS-1D and Sentinel-SLSTR benchmarks were evaluated on two architectures. These were a CPU system with two Xeon E5-2630-v3 processors, 20MB Cache, 64GB RAM, and 16 cores (32 hyper-threaded), and a GPU system with a TITAN-X (Pascal) GPU with 12GB DDR and 3840 GPU cores.

For both cases we report the classification performance (F1, accuracy and loss) and runtimes (training and inference time per sample). Wherever possible, we repeat the same across the different datasets.

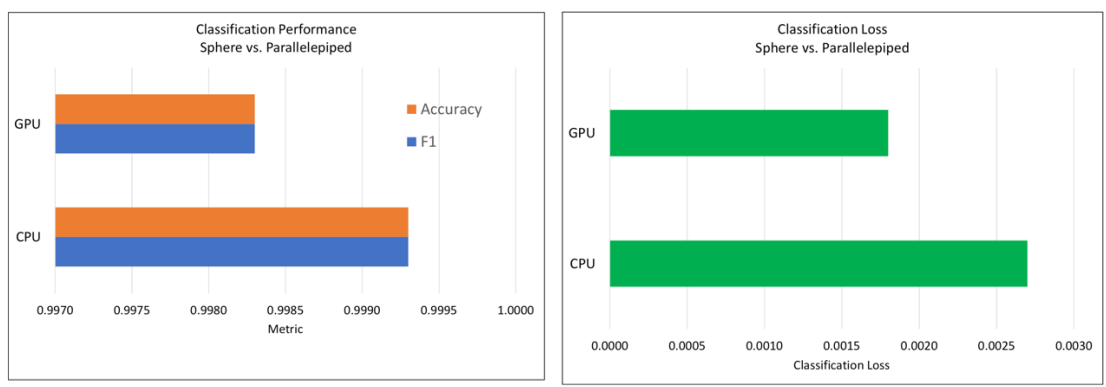

Fig. 13: Performance of the SAXS-baseline model on CPU and GPU systems. The figure shows the classification performance of the binary classification problem on the 1D profiles of mono-disperse shapes on two different datasets, on two different architectures.

In Figure 13, we show the classification performance of the SAXS-1D benchmark. The dataset has 20,000 1D profiles (with a 70:30 train:test split). For a simple baseline, it can be observed that different architectures yield different classification performance (both loss and accuracy). As the class divisions are even between the spheres and the parallelepiped, the F1 performance and the classification accuracy are the same here.

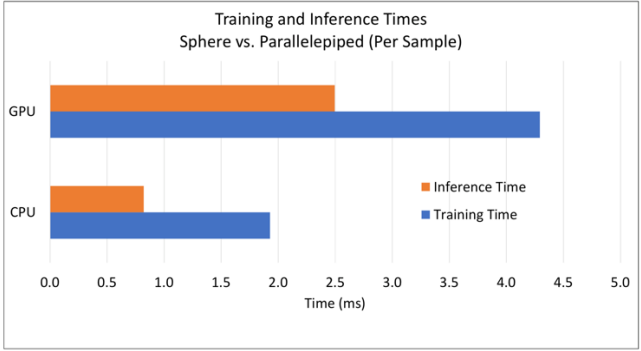

Fig.14. Training and inference time per sample across two datasets for the SAXS-1D benchmark.

We then show the overall training time and inference time performance for the same benchmark in Figure 14. The key observation here is that the inference time, as a percentage of overall training time, is different between two different architectures. More specifically, the inference time is 40% of the training time for the CPU architecture while for GPU it is 60%.

We show the classification and runtime performance for the Sentinel-SLSTR in Figure 15, using the dataset described above.

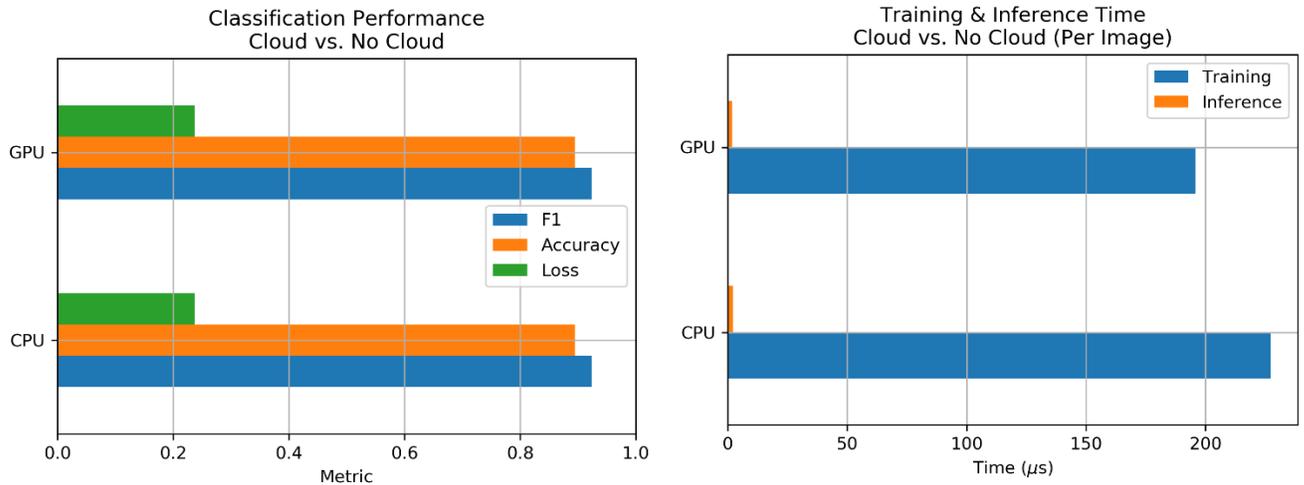

Figure 15. Classification and runtime performance for the Sentinel-SLSTR benchmark, where the classification task is to mark each pixel as 'Cloud' or 'Not Cloud'.

Given the baseline is with a single data set, we cannot draw any conclusions on the relationship between accuracy and dataset size. However, we observe that, as expected, the GPU architecture offers better training performance compared to the CPU platform.

## 5. Concluding Remarks

Deep learning is transforming many areas of computer science and underpinning the AI revolution that is happening around us. At the UK's Rutherford Appleton Laboratory, the large experimental facilities are now generating large volumes of increasingly complex data which will require new AI technologies to manage and interpret. In this paper we have given some examples of the opportunities for machine learning to play an important role both in the generation and analysis of some of these large datasets. In many areas of science, we are now seeing the emergence of a genuinely new 'Fourth Paradigm' of data-intensive scientific discovery [87,88]. For example, the combination of chemical databases, experimental data and detailed computer simulations is now leading to exciting new opportunities in materials science.

We have also introduced initial results on creating a scientific machine learning benchmark suite (SciML) covering a range of different scientific domains. Such a benchmark suite, based on scientific datasets of a significant scale and complexity, will enable scientists, computer scientists and data scientists to map out the applicability and limitations of deep learning neural networks and other machine learning algorithms applied to a range of real applications. Analysis of the SciML benchmark results will also reveal the strengths and weaknesses of the different computing platforms – from commercial Clouds and HPC systems to GPUs and FPGAs.

These benchmarks will also provide scientists with hands-on experience of using machine learning algorithms and environments on realistic-scale scientific datasets. In addition, the SciML benchmark suite will provide an important platform for research. One urgent research issue for scientists is the need to develop a disciplined framework for the uncertainty quantification (UQ) of deep learning algorithms. Another important issue is the need for transparency in understanding how such deep neural networks reach their conclusions. The robustness of deep learning predictions and their vulnerability to adversarial noise attacks also give genuine cause for concern. For applications in areas such as materials science and the life sciences, the challenge of incorporating physical, chemical or biological constraints into deep learning algorithms is an exciting topic for research. Despite these undoubted research challenges, the success of DeepMind's AlphaFold project has shown the effectiveness of deep learning for protein folding prediction. Could deep learning have a similarly transformative impact on other areas of data-intensive science?

**Acknowledgements**

The authors wish to thank Mark Basham; Tom Burnley, Martyn Winn and Jola Mirecka; Ben Davies and Dan Rolfe for their assistance in describing their work at the Diamond Light Source; at the Electron cryoEM Facility; and the OCTOPUS Laser Facility, respectively.

This work was supported by Wave 1 of The UKRI Strategic Priorities Fund under the EPSRC Grant EP/T001569/1, particularly the "AI for Science" theme within that grant & The Alan Turing Institute. We are grateful to James Hetherington, Oonagh McGee, Amit Mulji, Jon Rowe and Adrian Smith at the Turing Institute for their help and support.